\newif\iffulledition 
\fulleditiontrue 

\documentclass[\iffulledition nonacm ,\fi sigconf]{acmart}

\usepackage{booktabs}
\usepackage{xurl}

\copyrightyear{2023}
\acmYear{2023}
\setcopyright{rightsretained}
\acmConference[SAC '23]{The 38th ACM/SIGAPP Symposium on Applied Computing}{March 27--31, 2023}{Tallinn, Estonia}
\acmBooktitle{The 38th ACM/SIGAPP Symposium on Applied Computing (SAC '23), March 27--31, 2023, Tallinn, Estonia}
\acmDOI{10.1145/3555776.3577805}
\acmISBN{978-1-4503-9517-5/23/03}

\begin{document}

\title[Cyber Security and Online Safety Education for UK Schools]{Cyber Security and Online Safety Education for Schools in the UK: Looking through the Lens of Twitter Data}
\iffulledition
\titlenote{This is the full edition of a 4-page poster paper published in the Proceedings of the 38th ACM/SIGAPP Symposium on Applied Computing (SAC '23), which can be accessed via the following DOI link: \url{https://doi.org/10.1145/3555776.3577805}.}
\titlenote{Please cite this paper as follows: Jamie Knott, Haiyue Yuan, Matthew Boakes and Shujun Li, ``Cyber Security and Online Safety Education for Schools in the UK: Looking through the Lens of Twitter Data,'' arXiv:2212.13742 [cs.CY], 28 December 2022, \url{https://doi.org/10.48550/arXiv.2212.13742}, a shorter edition was published in Proceedings of the 38th ACM/SIGAPP Symposium on Applied Computing (SAC '23), ACM, 2023, \url{https://doi.org/10.1145/3555776.3577805}.}
\fi

\author{Jamie Knott, Haiyue Yuan, Matthew Boakes and Shujun Li}
\iffulledition
\authornote{Authors can be contacted via \{jamieknott7, h.yuan-221, m.j.boakes, s.j.li\}@kent.ac.uk.}
\fi
\affiliation{%
\institution{Institute of Cyber Security for Society (iCSS) \& School of Computing, University of Kent, Canterbury, UK}
}
\renewcommand{\shortauthors}{Knott et al.}

\begin{abstract}
In recent years, digital technologies have grown in many ways. As a result, many school-aged children have been exposed to the digital world a lot. Children are using more digital technologies, so schools need to teach kids more about cyber security and online safety. Because of this, there are now more school programmes and projects that teach students about cyber security and online safety and help them learn and improve their skills. Still, despite many programmes and projects, there is not much proof of how many schools have taken part and helped spread the word about them. This work shows how we can learn about the size and scope of cyber security and online safety education in schools in the UK, a country with a very active and advanced cyber security education profile, using nearly 200k public tweets from over 15k schools. By using simple techniques like descriptive statistics and visualisation as well as advanced natural language processing (NLP) techniques like sentiment analysis and topic modelling, we show some new findings and insights about how UK schools as a sector have been doing on Twitter with their cyber security and online safety education activities. Our work has led to a range of large-scale and real-world evidence that can help inform people and organisations interested in cyber security and teaching online safety in schools.
\end{abstract}




\maketitle

\section{Introduction}

\iffulledition Research has repeatedly shown that increasingly higher numbers of young people are using technology and the internet. The time spent by adolescents on these technological devices has also seen major increases, especially over the past few years. \fi In 2019, the United Nations International Children's Emergency Fund (UNICEF) reported results from a survey involving more than 14,000 internet-using children across 11 countries in four continents (Europe, South America, Africa, and Asia). According to the survey results, the average internet usage was two hours per day during the week and roughly doubled the time on a weekend day~\cite{Stalker-P2019}. \iffulledition In addition to the worldwide statistics showing how frequently children and young people are exposed to the internet, there are also national ones giving more detailed statistics~\cite{statista2021}. For instance, according to statistics provided by Statista~\cite{statista2021}, the time spent on the internet per week increased steadily from 2014 to 2018, for different age groups of children living in the UK, even for those as young as 3–4 years old (who spent 6.6 hours in 2014, but this increased to 8.9 hours in 2018). \fi Even though most people agree that technology and the internet are suitable for kids and teens, there are also risks to their safety when they use them more. There have already been some worrying statistics published~\cite{ons_children_online_behaviour_2021}, e.g., around 1 in 6 children aged 10 to 15 years had spoken to a stranger online in 2020, \iffulledition 5\% of children aged 10 to 15 years had physically met with someone they had only spoken to online, \fi and around 1 in 10 children aged 13 to 15 years had experienced receiving sexual messages. Therefore, to protect children and young people who know how to protect themselves online, equipping them with relevant knowledge and skills in cyber security and online safety becomes very important.

Many cyber security and online safety education programmes and initiatives targeting children and young people have been launched to meet the above mentioned needs. Some nations have started including relevant content in their national curricula or guidelines\iffulledition~\cite{Waldock2022GFCE}\fi. Despite all the educational activities on cyber security and online safety education, there is still little effort on how such activities are received, and a particular area with even less evidence is \emph{to what extent schools (i.e., pre-university educational institutions) have been actively engaging and publicly promoting such activities}. This paper tries to fill this gap by exploring if and how Twitter data can be used to infer insights about schools' engagement in cyber security and online safety education, using the UK as an example country and nearly 200k public tweets from over 15k schools. \iffulledition Our work leverages simple data analysis techniques such as descriptive statistics and visualisation and more advanced ones such as topical modelling and sentiment analysis based on natural language processing (NLP) techniques. \fi Our work provides positive evidence about the data-driven approach to the research question. It produces valuable insights for researchers, practitioners, and policymakers interested in cyber security and online safety education for schools.
\iffulledition
The main findings include the following:
\begin{itemize}
\item There are many orgnisations have launched dedicated programmes and initiatives for improving pupils' cyber security awareness and offering cyber security education to pupils.

\item Schools in the UK have been actively engaging and promoting related activities through the use of their Twitter accounts.

\item Both maintained schools and academies have been linked with more cyber security and online safety organisations and initiatives than other school types (i.e., private school, special school and college).

\item The sentiment analysis concludes that the majority of tweets posted by schools have positive sentiment, suggesting that schools in the UK are generally having positive attitudes and opinions towards the cyber security and online safety education.

\item The topic modelling of large scale tweets data highlighted the core topics including safety, cyber security, personal, delivery method, information medium, online behaviour and activity, which reflects the main elements of the cyber security and online safety education in the UK. 
\end{itemize}
\fi

\iffulledition The rest of the paper is organised as follows. First, Section~\ref {sec:related_work} presents the relevant projects and most related work. Section~\ref{sec:methodology} presents some relevant background for the school education system and the status of cyber security and online safety education in the UK and then outlines the methodology of our study. Then, Section~\ref{sec:data_collection} describes the process of preparing and collecting data using publicly available sources for this study, followed by presenting the data analysis and results in Section~\ref{sec:data_analysis}. Finally, the last section concludes this paper with a discussion and future work. \fi

\section{Related Work}
\label{sec:related_work}

Children and teens spend much time on the internet, which can expose them to online risks. This exposure has made it more critical for schools to teach cyber security and online safety. \citet{Rahman-N2020} conducted a systematic review to state that the critical reason for having cyber security education in schools is to educate children to become aware of the associated risks of using online services such as social media, chatting, and gaming. Furthermore, \citet{Macaulay-P2020} was surveyed to evaluate the impact of children's subjective and objective knowledge on their perception and attitudes towards online safety education, concluding that online safety education is essential, especially for children lacking awareness and knowledge. \iffulledition Moreover, based on a review of related policy, educational, and implementation landscapes on cyber security and online safety education and several interviews with key stakeholders, \citet{Waldock2022GFCE} established that providing children and young people with cyber security and online safety education in pre-university settings is crucial. Also, some problems need to be fixed, such as the fact that this kind of education is scattered.\fi

In addition, much research has been conducted to understand the current practice and status of introducing cyber security and online safety education in schools and investigate the effectiveness of related programmes and initiatives for pupils regarding their perceptions and attitudes. \iffulledition Here, we briefly review some related work. \citet{Ranguelov-S2010} used a questionnaire coordinated by the European Education and Culture Executive Agency to discover that education authorities in Europe in collaboration with public and private organisations can integrate cyber security and online safety in various subjects in school curriculum. Regarding cyber security and online safety education, a 2022 report~\cite{Waldock2022GFCE} gives a very recent and comprehensive summary based on findings from 13 countries on five continents (Europe, North America, Africa, Asia, and Australasia). \fi It finds that the current practice of embedding cyber security and online safety content into the pre-university education curriculum is either by adding the educational content to technical subjects such as computing and computer science or by adding the content to a range of non-technological subjects. Much research has not been done on how well real-world cyber security and online safety education programmes work for kids. However, much of the research looked at the effectiveness of various tools for cyber security education at the pre-university level~\cite{Zhang-KL2021}.

\iffulledition
For example, a 2021 survey \cite{Zhang-KL2021} reviewed multimedia tools, including games, comics, films, and tabletop for cyber security awareness and education and observed that only 30\% of all tools reviewed from 2000 to 2019 were evaluated to assess their effectiveness and usability. Because of this, the authors stressed the need for structured and thorough ways to measure and validate how well cyber security education tools work. Similarly, another 2022 systematic literature review conducted by~\citet{Saglam-R.B2022} discovered that using computer-based methods such as virtual environments, interactive books, games, and digital storytelling could improve the quality of cyber security education. However, such methods are still being used to evaluate real-world cyber security and online safety education initiatives.
\fi

\iffulledition Even though there is much research on cyber security and online safety education for schools, one thing that stands out is that most of the previous studies were small-scale empirical studies that relied on the self-reported perceptions of recruited human participants, such as surveys, interviews, and observational studies of educational events. \fi We have not seen as much research that looks at large amounts of data in the real world to study cyber security and online safety education in the real world. To the best of our knowledge, the only work similar to ours reported in this paper was done by~\citet{Zenebe-A2018}. Their research aimed to find patterns and insights about how people in the USA think about cyber security education in general. \iffulledition They collected 1,387 tweets with the hashtags \#cybersecurity and \#education between 2015 and 2017. Descriptive statistics and sentiment analysis were used to address their research questions.  \fi While comparing our work to theirs, we focused on a different research question: how much have schools in the UK promoted cyber security and online safety education to the public? Our analysis is much more advanced regarding the size and quality of data. We used nearly 200,000 tweets from more than 15,000 verified school accounts, which were chosen from lists of all UK schools kept by the government.

\section{Background and Methodology}
\label{sec:methodology}

To study schools' engagement and public promotion of cyber security and online safety education, we decided to use public tweets of verified schools' accounts because we observed that many schools have an active presence on Twitter and the Twitter API allowed us to gather timelines of a Twitter account quickly. In order to get a list of verified school accounts, we felt the need to focus on a single country so that we could rely on that country's education authorities to obtain official information about recognised schools. 

\iffulledition
Out of all the countries, we decided to choose the UK because it has the most active cyber security and online safety educational activities, according to some recent reports~\cite{Waldock2022GFCE, Dcms_report2022}. Note that our methodology is general, so it does not depend on this specific choice of country.
\fi

\iffulledition
At the highest level, the UK has several national strategies and policies promoting the importance of cyber security for all sectors, including cyber security education and skill development, e.g., the \emph{National Cyber Strategy 2022} produced by the Cabinet Office, the 2019 \emph{Online Media Literacy Strategy} launched by the DCMS, and the 2021 policy document ``\emph{Keeping children safe in education}'' published by the DfE~\cite[Section~4.2]{Waldock2022GFCE}. In terms of school education, each of the United Kingdom's four countries – England, Wales, Scotland, and Northern Ireland – has its national curriculum, and cyber security and online safety content have been introduced in all four national curricula and is also reflected in various pre-university qualifications assessment standards~\cite[Section~4.4]{Waldock2022GFCE}. 

Accordingly, most schools across the UK offer compulsory and optional cyber security teaching to pupils~\cite[Table~9]{Waldock2022GFCE}. In addition, much extracurricular cyber security and online safety education and awareness programmes and initiatives have been organised by different public and private sector organisations, e.g., \emph{CyberFirst} of the National Cyber Security Centre (NCSC), \emph{Cyber Discovery} of the DCMS, \emph{Cyber Choices} of the National Crime Agency (NCA), and the UK Safer Internet Centre run by a partnership of three private sector organisations~\cite[Section~4.4]{Waldock2022GFCE}, and many other activities run by non-governmental bodies~\cite{Dcms_report2022}.
\fi

With the UK chosen as the country of interest, we used a five-step process to do our work:
1) gathering official information of all schools in the UK; 2) using the school information to automatically gather several Twitter accounts that may belong to a school; 3) using a semi-automatic process to identify Twitter accounts that belong to a school; 4) collecting timelines of all verified school Twitter accounts and using a data cleansing process to get tweets related to cyber security and online safety education; and 5) applying different data analytic techniques to investigate how much schools have publicly engaged and publicly promoted cyber security and online safety education on Twitter to discover valuable insights for relevant stakeholders. It is worth noting that the first four stages are all about data collection, and the last one is about the analysis of collected data. In the following section, we give details of the four data collection stages, and Section~\ref{sec:data_analysis} shows how we conducted our data analysis and critical findings.

\iffulledition
\begin{figure*}[!htb]
\centering
\includegraphics[width=0.7\linewidth]{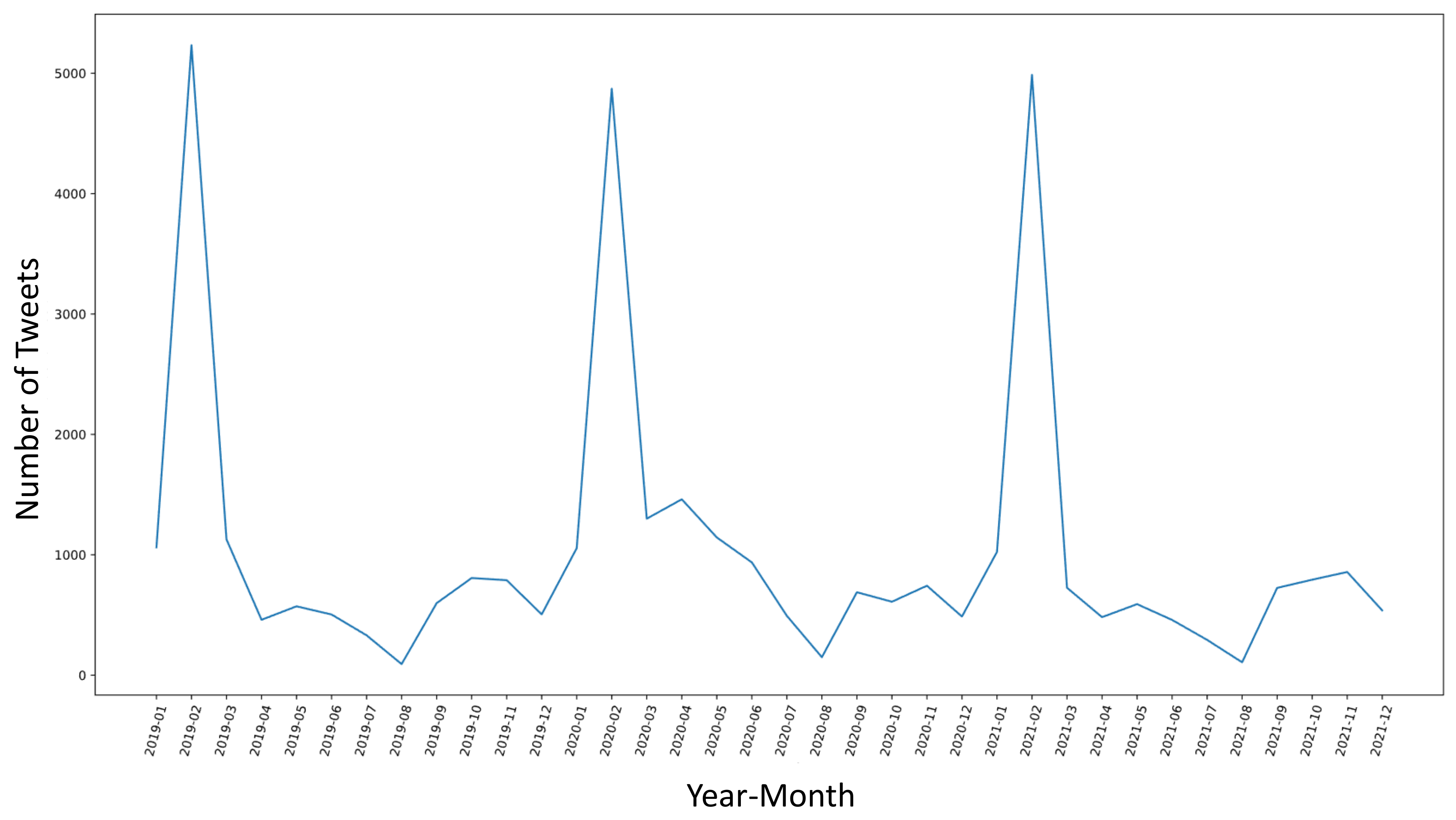}
\caption{The number of tweets across all UK schools from 2019 to March 2021}
\label{fig:tweets_activity}
\end{figure*}
\fi

\section{Data Collection}
\label{sec:data_collection}

\iffulledition
We used Google search and checked the websites of educational authorities in the UK and the four countries within the UK to identify official lists of schools in the UK. This search led us to four primary sources: an online portal of the ONS covering schools in England~\cite{ONS2022schools}, a list of Welsh schools published by the Welsh Government~\cite{welsh_school_list_2022}, a 2014 list of Scottish schools published by the Scottish Government~\cite{scottish_school_list_2014}, and an online portal of the Department of Education for Northern Ireland~\cite{DENI2022schools}. Each source provides a school dataset with helpful information, such as the school's name and location. See Table~\ref{tab:stats_schools_accounts} for the number of schools per country and the whole UK included in the four datasets.
\fi

\iffulledition
\subsection{Collecting Information about Schools}

We used Google search and checked the websites of educational authorities in the UK and the four countries within the UK to identify official lists of schools in the UK. This search led us to four primary sources: an online portal of the ONS covering schools in England~\cite{ONS2022schools}, a list of Welsh schools published by the Welsh Government~\cite{welsh_school_list_2022}, a 2014 list of Scottish schools published by the Scottish Government~\cite{scottish_school_list_2014}, and an online portal of the Department of Education for Northern Ireland~\cite{DENI2022schools}. Each source provides a school dataset with helpful information, such as the school's name and location. See Table~\ref{tab:stats_schools_accounts} for the number of schools per country and the whole UK included in the four datasets.

\begin{table}[!htb]
\centering
\caption{Descriptive statistics on schools and verified school Twitter accounts (E = England, S = Scotland, W = Wales, NI = Northern Ireland)}
\label{tab:stats_schools_accounts}
\begin{tabular}{*{6}{c}}
\toprule
& E & S & W & NI & UK\\
\midrule
Schools & 25,924 & 2,461 & 1,472 & 1,146 & 31,003\\
Twitter accounts & 12,541 & 1,033 & 1,013 & 422 & 15,010\\
Percentage & 48\% & 42\% & 69\% & 37\% & 48\%\\
\bottomrule
\end{tabular}
\end{table}

\fi

\iffulledition
\subsection{Collecting Candidate Twitter Accounts}

Given the name and location of a school, a two-step search algorithm was made to find the Twitter account handles automatically:
\begin{itemize}
\item \textbf{Step~1:} A school's name and location information are used as keywords to search on Google, leading to a list of search results. Next, the Twitter handle is taken from the page that links to a Twitter account by looking at the URLs of the returned results. This step was automated using a web browser automation tool Selenium\footnote{\url{https://www.selenium.dev/}}.

\item \textbf{Step~2:} Each Twitter handle obtained from Step~1 is searched using the Twitter Search API to obtain its profile data, such as name, location, description, and URL. Next, a similarity score is made for each Twitter handle based on tokenization and Levenshtein distance. This score is found by comparing the information in each Twitter handle's profile to the information in the official dataset about the target school. Finally, among all the Twitter accounts given a school's name and location information, the Twitter profile account with the highest similarity score is selected as the school's Twitter account.
\end{itemize}

We used Google search as Step 1 because we found that using the Twitter Search API alone does not give good results. However, after running the above algorithm on all of the schools in our dataset, we found the Twitter accounts of 19,811 unique candidates to check further.

\fi

\iffulledition
\subsection{Verifying Twitter Accounts of Schools}

All candidate Twitter accounts were grouped into four groups according to the country the school belongs to (i.e., England, Wales, Scotland, and Northern Ireland). Semi-automatic validation was used to eliminate Twitter accounts that are not officially linked to a school. When we looked at some real school Twitter accounts, we saw that their profile URLs all pointed to the school's website. This discovery is not a big surprise since schools promote themselves on Twitter and should have a reason to promote their website there. Based on this, we made a simple but effective three-step rule-based algorithm for checking the URL of a candidate's Twitter account to confirm it:
\begin{itemize}
\item \textbf{Step~1:} If the URL's domain name has ``.sch'' as the second-level domain name and ``.uk'' as the top-level domain name, the account is labelled as genuine; otherwise, go to Step~2. This step is based on the observation that most UK schools choose to register a ``.sch.uk'' domain name\footnote{\url{https://nominet.uk/wp-content/uploads/2020/01/Schools_Domain_Name_Rules_2020.pdf}}.

\item \textbf{Step~2:} Search for the indicative keyword ``school'', ``academy'' or ``college'' in the content of the \texttt{<title>} element of the web page the URL points to. If such a keyword exists, label the account as genuine. Otherwise, go to Step~3.

\item \textbf{Step~3:} Count the frequencies of three school-related keywords: ``school'', ``academy'' and ``college'' in the body of the web page the URL directs. If any frequency exceeds a threshold, label the account as genuine. Otherwise, label the account as ``unverified''. The threshold was set to 5 after manually inspecting some randomly selected accounts.
\end{itemize}

Before deploying the algorithm to all unverified Twitter accounts, we tested it with a smaller dataset of 100 randomly sampled Twitter accounts with URLs in their profiles. We then manually validated the results, concluding that the algorithm could successfully verify school Twitter profiles with 100\% accuracy.

All 19,811 Twitter accounts were checked with this algorithm, and 12,249 were confirmed as official school Twitter accounts. The rest of the unverified Twitter accounts were checked by hand by the first author of this paper. Next, the first author manually verified them by examining the Twitter profiles, resulting in an additional 2,761 verified Twitter accounts. So in total, we got 15,010 verified Twitter accounts. Table~\ref{tab:stats_schools_accounts} shows how the verified school Twitter accounts vary across the four UK countries. As a whole, nearly half of all schools have a verified Twitter account, and the rate is much higher for Wales (69\%) but the lowest for Northern Ireland (37\%).

\fi

\iffulledition
\subsection{Collecting and Cleansing Twitter Data}

After obtaining the verified Twitter accounts, the Python library snscrape~\cite{snscrape} was used to retrieve all tweets posted by these accounts from 2009 to March 2022, resulting in 20,617,709 tweets. These raw tweets data are a mixture of different data formats such as texts, images, videos, and URLs. A further data cleansing step was performed following the following three steps to ensure the data is adequate for the follow-up NLP-based analysis: 1) removing non-textual data from all tweets; 2) using several keywords extracted from the cyber security and online safety-related content in the national curricula of the four UK countries to narrow down relevant tweets, and 3) filtering out tweets written in Welsh.

By applying the steps above to the whole raw dataset, we got 193,424 tweets ready for the last stage of our data analysis work. Finally, to clean the data for follow-up NLP operations, we used the Python-based NLP library NLTK~\cite{nltk} to clean the data by removing stop-words, URLs, punctuation marks, email addresses, Twitter screen names, and other non-textual data such as symbols, emoticons, and icon emojis.

\fi

\section{Data Analysis and Results}
\label{sec:data_analysis}

Different techniques were utilised for analysing the cleaned data. Firstly, an exploratory data analysis (EDA) approach was applied to explore and visualise the data to get some initial insights. Then, NLP-based techniques such as sentiment analysis and topic modelling were applied to analyse the data further for more in-depth insights regarding how the schools covered have engaged with and publicly promoted cyber security and online safety education on Twitter.

\begin{table*}[!htb]
\centering
\caption{Some example organisations and associated initiatives with keywords we used}
\label{tab:keywords}
\begin{tabular}{cccc}
\toprule
Organisation & Keywords & Associated Initiatives  & Keywords\\
\midrule
NCSC & NCSCgov, \#NSCS, @NCSC & Cyber Aware & CyberAware,  cyberawaregov\\
DCMS & \#DCMS, @DCMS Cyber Discovery &  CyberDiscovery & CyberDiscUK\\
BCS & BCS, @BCS, \#BCS & Computing at School & CAS, CASInspire, CompAtSch, CASPLYM22\\
\bottomrule
\end{tabular}
\end{table*}

\subsection{Exploratory Data Analysis}

\subsubsection{Twitter Activity}
\label{sec:twitter_activity}

\iffulledition
The visualisation of Twitter activity amongst schools from 2019 to 2021 is presented in Figure~\ref{fig:tweets_activity}, where the y-axis denotes the number of tweets posted and the x-axis represents the timeline. The three dominating peaks for the activity appear to be Tuesday of the second week in February of every year, which were identified as the Safer Internet Days (SIDs, \url{https://www.saferinternetday.org/}): 5th February 2019, 11th February 2020, and 9th February 2021. This discovery indicates that the most successful cyber security and online safety education initiative across UK schools is likely the SID.
\fi
The SID is an initiative launched by the EU SafeBorders project in 2005 and has grown significantly worldwide with the participation of approximately 200 countries and regions. While it is not surprising that SID is the most successful cyber security and online safety education initiative across UK schools, this is the first time that such large-scale evidence has been produced for this fact. Therefore, extending our work to see how schools in other countries and regions have been engaged with and publicly promoting this initiative could potentially offer valuable insights in the global context.

\subsubsection{Mentions of Organisations \& Initiatives}

As mentioned in Section~\ref{sec:methodology}, many organisations in the UK have been actively running other cyber security and online safety education initiatives. Therefore, we are interested in seeing how much schools in the UK have engaged with and publicly promoted a more comprehensive range of organisations and initiatives on Twitter. To this end, we manually populated a list of keywords (e.g., organisations' or initiatives' names, acronyms, alternative names, Twitter handles, and hashtags) associated with key organisations and initiatives in the UK (see Table~\ref{tab:keywords} for some examples). In addition, we count the number of schools whose tweets mention each keyword at least once.

\iffulledition
\begin{figure}[!htb]
\centering
\includegraphics[width=0.9\linewidth]{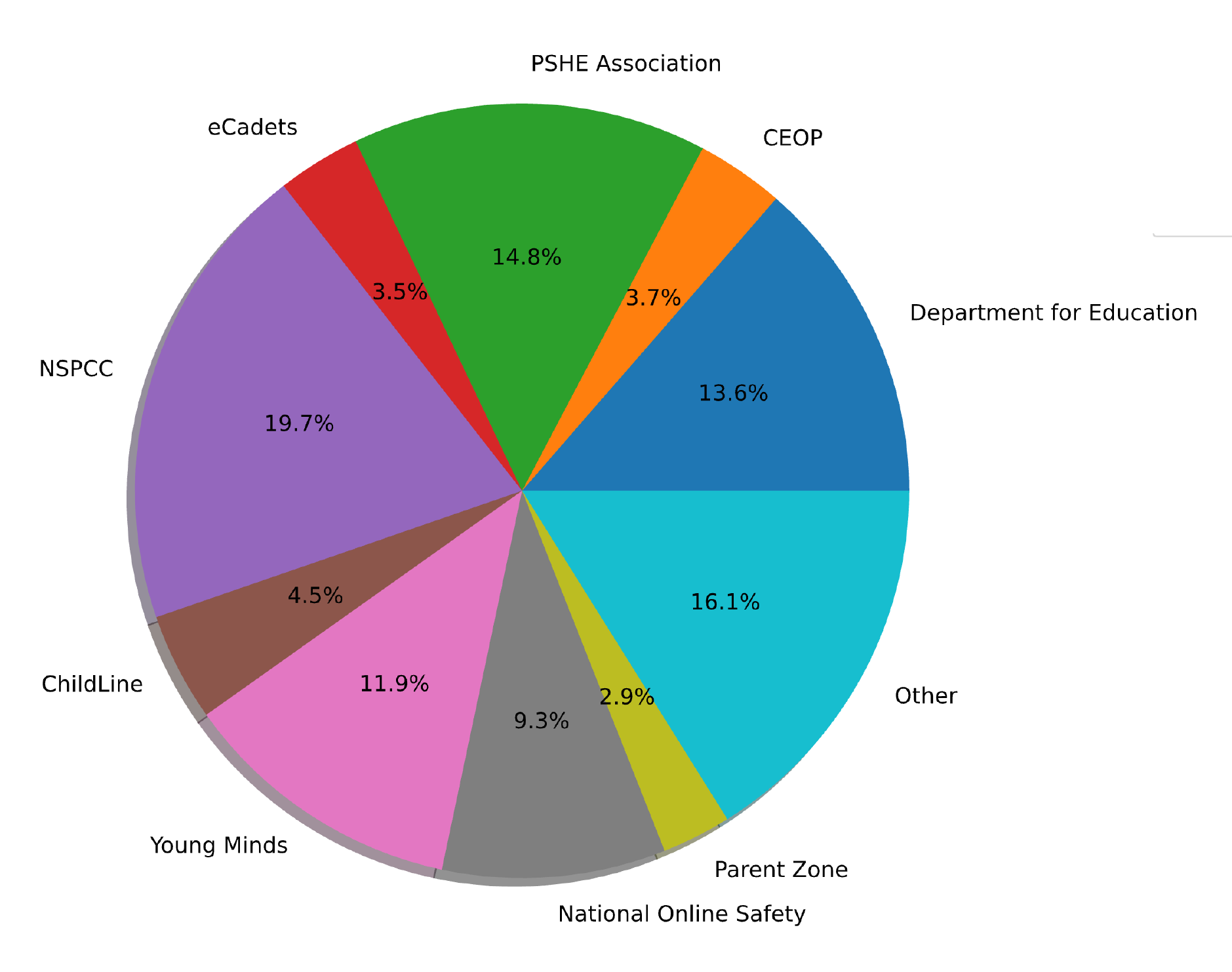}
\caption{Most-mentioned organisations by UK schools regarding cyber security and online safety education}
\label{fig:most_mentioned_organisations}
\end{figure}
\fi

By searching all tweets using the generated keywords, we summarised the mentioned frequency of each keyword. \iffulledition As shown in Figure~\ref{fig:most_mentioned_organisations}, the pie chart illustrates \fi The organisations with mentions by at least 700 schools include the National Society for the Prevention of Cruelty to Children (NSPCC); Personal, Social, Health and Economic (PSHE) Association; Childline; Young Minds; National Online Safety (NOS); and UK Government's Department for Education (DfE). Among all schools, the organisation with the most mentions is the NSPCC (19.7\%), closely followed by the PSHE Association (14.8\%) and the DfE (13.6\%). \iffulledition The ``Other'' sector in the pie chart represents organisations mentioned by less than 700 schools in the UK. \fi In addition, \iffulledition as illustrated in Figure~\ref{fig:most_mentioned_initiatives}, \fi we conducted a similar analysis using the keywords associated with selected initiatives and compared initiatives mentioned in at least 100 schools. UK schools mainly mentioned the Safer Internet Day initiative, echoing the observation of Twitter activity shown in \iffulledition Figure~\ref{fig:tweets_activity}. \fi

\iffulledition
In addition, as illustrated in Figure~\ref{fig:most_mentioned_initiatives}, we conducted a similar analysis using keywords associated with selected initiatives and compared initiatives mentioned by at least 100 schools. UK schools mentioned mainly the SID initiative, echoing the observation of the Twitter activity shown in Figure~\ref{fig:tweets_activity}.
\fi
Furthermore, the second most-mentioned initiative is ``ThinkUKnow'', a programme organised by the Child Exploitation and Online Protection Command (CEOP) education team within the NCA. The third most-mentioned initiative among all schools is ``Wake Up Wednesday'', an online safety programme run by the NOS. Moreover, it is worth noting that among all the initiatives mentioned by fewer than 100 schools, the dominating initiative is Cyber Discovery, a programme of the UK Government's Department for Digital, Culture, Media \& Sport (DCMS).

\iffulledition
\begin{figure}[!htb]
\centering
\includegraphics[width=\linewidth]{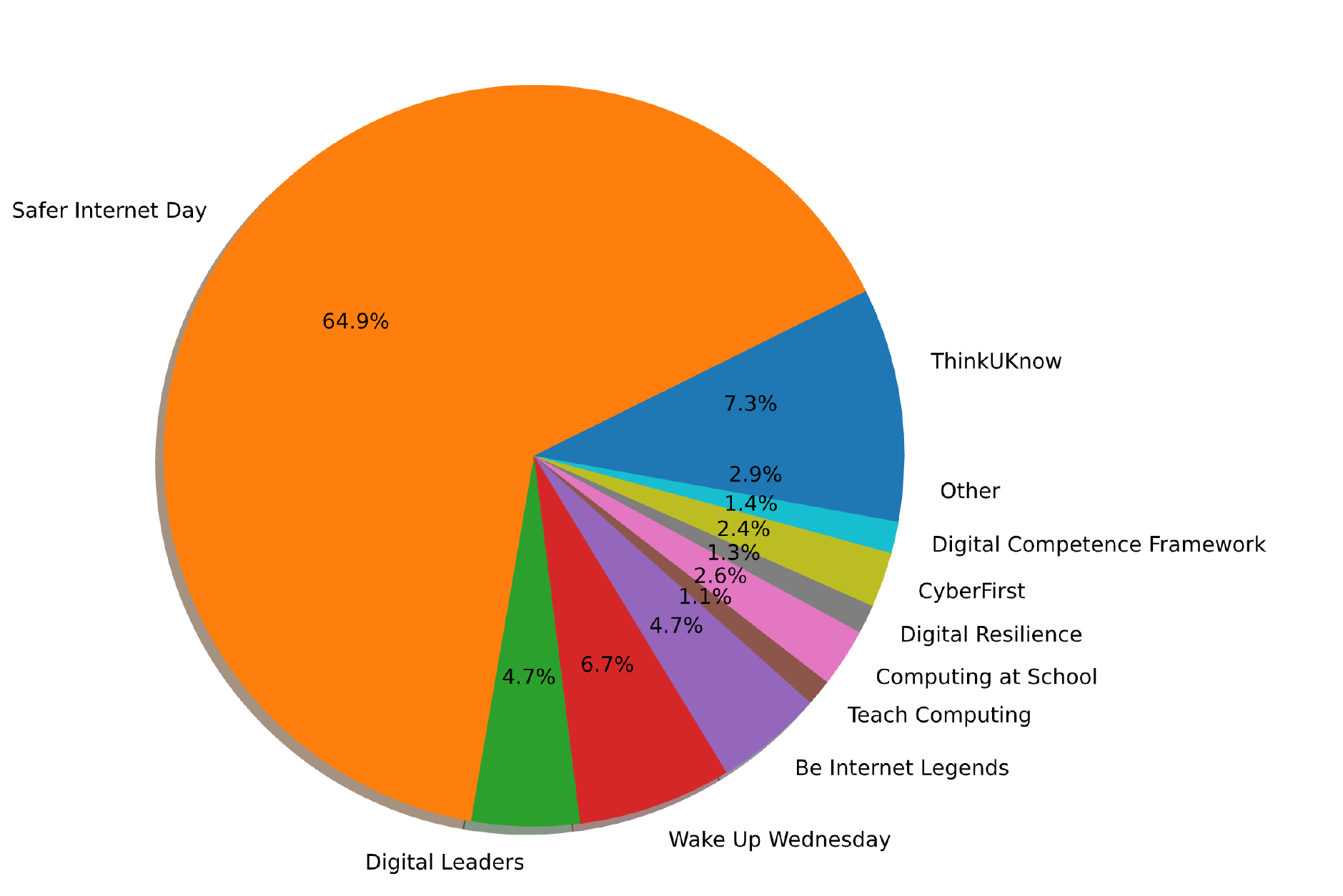}
\caption{Most mentioned initiatives in the UK}
\label{fig:most_mentioned_initiatives}
\end{figure}

\begin{figure*}[!htb]
\centering
\includegraphics[width=0.9\linewidth]{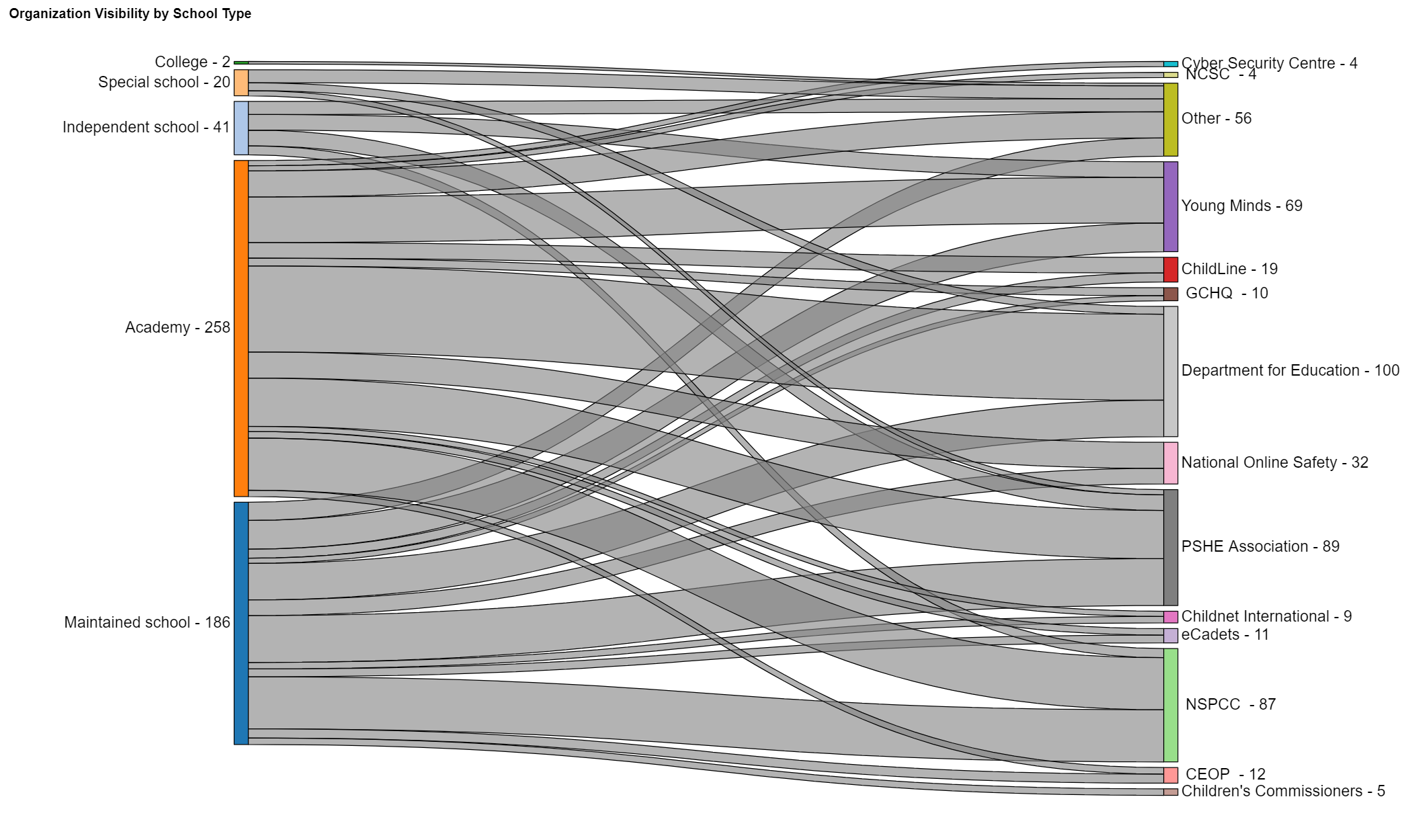}
\caption{Organisations mentioned by schools of different types (the larger a stream is, the more schools mentioned the corresponding organisation)}
\label{fig:sankey}
\end{figure*}
\fi

\subsubsection{Engagement Diversity by School Type}

We also wanted to know if cyber security and online safety education have been taught and promoted differently by different types of schools. In the four UK countries, there are different types of schools. Using England as an example, there are five main types of schools according to the dataset of English schools we used: (state-)maintained schools, independent (private) schools, academies, special schools, and colleges\iffulledition~\cite{schooltype}\fi. A sample of 1,000 verified Twitter accounts of English schools were selected randomly, and a Sankey diagram was produced to show how the different types of schools mentioned different organisations, as illustrated in Figure~\ref{fig:sankey}.

The results revealed significantly different engagement patterns. Both academies and maintained schools have a diverse engagement profile: they have been actively engaging with 13 and 12 organisations, respectively. However, independent schools have a much less active profile by engaging with only four organisations. One noticeable pattern is that independent schools did not post any tweets that mentioned the DfE, which may be explained by the fact that independent schools do not rely on the resources of the DfE, nor do they have to follow the national curriculum. Given the richer resources at independent schools, we found their lack of engagement worrying, and more work should be done to motivate them to do more. The remaining two types of schools have an even less active profile: special schools only engaged with three organisations, and colleges did not engage with any leading organisations shown in Figure~\ref{fig:sankey}. Such a noticeable discrepancy may be rooted in the different goals and interests of different types of schools. \iffulledition
However, it may also echo the fragmented nature of cyber security and online safety education in the UK and elsewhere, as reported in~\cite{Waldock2022GFCE}.\fi

\subsection{NLP-based Content Analysis}

In addition to what has been described above, we also applied two NLP-based techniques, sentiment analysis and topic modelling, to analyse the Twitter data to understand more about the content of the school's Twitter accounts.

\subsubsection{Sentiment Analysis}

The essential task of sentiment analysis is to classify if the sentiment status of a given text is positive, negative or neutral. Such an analysis helps analyse the emotional status of the author of a given text. It, therefore, has been widely used for analysing user-generated textual data such as tweets in different applications~\iffulledition\cite{Agarwal-A2011, Kouloumpis-E2021, Martinez-CE2014, Drus-Z2019}\fi. In this study, an off-the-shelf sentiment analysis package provided by NLTK\iffulledition\footnote{\url{https://www.nltk.org/api/nltk.sentiment.html}} \fi was used to conduct sentiment analysis to quantify the sentiments of tweets data to infer the schools' opinions and attitudes towards cyber security education. By analysing all tweets posted by schools, 82.2\% of all tweets are classified as positive, whereas 12.8\% and 5\% of tweets are classified as negative and neutral, respectively.

It is not surprising to see a majority of the posted tweets with positive sentiment, as many schools' Twitter accounts are professionally managed and maintained, so it is less likely for them to publicly express negative opinions on other organisations and initiatives with a good purpose.
We manually examined some negative tweets and noticed that some are misclassified because of the use of some negative-meaning words such as `cyber bullying'. This misclassification suggests that the rate of negative sentiment is likely over-estimated, so even more tweets should be considered positive or neutral.

\iffulledition
\begin{table*}[!h]
\centering
\caption{Top 5 terms for the identified 6 topics}
\label{tab:topics}
\begin{tabular}{*{6}{c}}
\toprule
\textbf{Topic label} & \textbf{Term 1} & \textbf{Term 2} & \textbf{Term 3} & \textbf{Term 4} & \textbf{Term 5}\\
\midrule
\emph{Safety}  & esafety & internet & digital literacy & digital footprint & strangers \\
\emph{Cyber security} & cyber & passwords & coding & scams/phishing & permissions \\
\emph{Personal data} & identity & data & details & information & health \\
\emph{Delivery methods} & workshops & assemblies & sessions & presentation & lessons\\
\emph{Delivery formats} & posters & guides & books & newsletters & websites \\
\emph{Online behaviour/Activity} & cyberbullying & sexting & messaging & gaming & games\\
\bottomrule
\end{tabular}
\end{table*}
\fi

\subsubsection{Topic Modelling}

Topic modelling is an NLP technique for identifying the main topics in a collection of texts, which could help identify hidden semantic structures in a text body. It has been used frequently for analysing Twitter data in different contexts~\cite{Ostrowski-DA2015,Negara-ES2019}. In this study, we used the latent Dirichlet allocation (LDA) provided by Gensim~\cite{gensim}, a popular topic modelling and NLP Python library, to analyse our Twitter data, aiming to identify different popular topics UK schools discussed regarding cyber security and online safety education on Twitter.

One of the critical parameters of using the LDA is to specify the number of topics. To determine this parameter, we manually examined and compared the results of topics and their associated terms generated by the LDA model using different numbers of topics. Then we decided to use six as the best number because it gives the most meaningful set of topics. \iffulledition As shown in Table~\ref{tab:topics}, six distinct topics are extracted with manually added labels. \fi These identified topics are closely interconnected and reflect the core concepts and the general scope of cyber security and online safety education. \emph{Topic 1: Safety} and \emph{Topic 2: Cyber security} are important to protect \emph{Topic 3: Personal data} including identify, data, and health-related information. \emph{Topic 4: Delivery methods} mainly represents how to deliver cyber security and online safety education programmes/initiatives to schools, where workshops, assemblies, and talks are amongst the most popular physical delivery methods. Many tweets aim to market or showcase cyber security and online safety events organised at school, especially those hosted by external organisations and partners such as the NOS or the NCA. \emph{Topic 6: Delivery formats} is more related to the different formats of content delivery, often in the form of online safety posters, websites/links, guides, and newsletters. Numerous tweets aimed to share these resources with the relevant stakeholders and followers (such as parents and pupils).

\iffulledition
To further complement the topic modelling results, we summarised the frequencies of essential keywords extracted from the national curricula in the UK. As depicted in Figure~\ref{fig:curriculum}, `Online safety' is, by far, identified as the most mentioned keyword (59.4\%), followed by `passwords' (8.6\%), `privacy' (6.7\%) and `cyber bullying' (7\%). Again, these are primarily aligned with the findings from the topic modelling results and our general understanding of what should be considered by schools.

\begin{figure}[!htb]
\centering
\includegraphics[width=\linewidth]{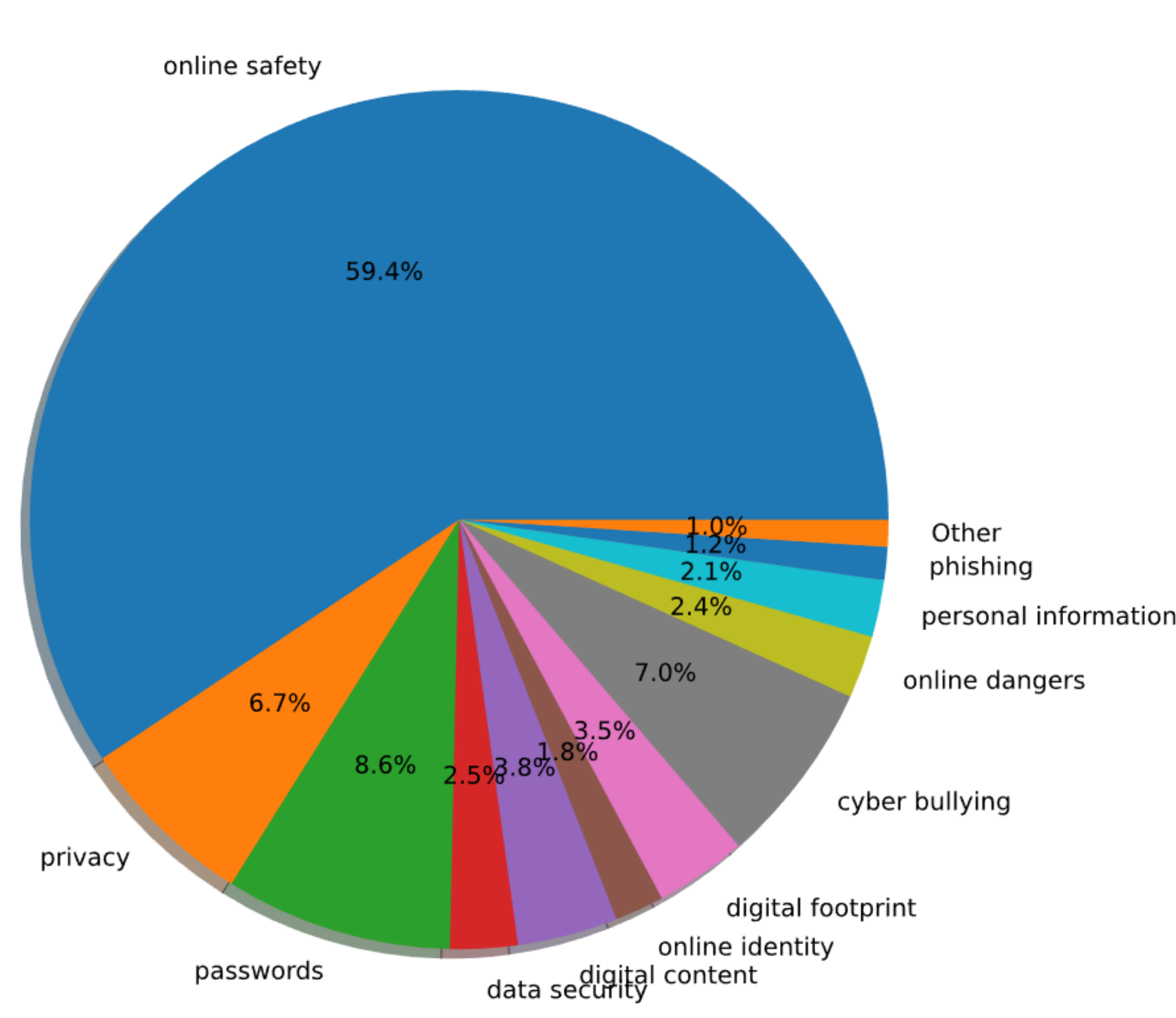}
\caption{Most popular cyber security and online safety curriculum topics in the UK}
\label{fig:curriculum}
\end{figure}
\fi

\section{Conclusion}

Via a data-driven analysis of nearly 200k tweets from over 15k Twitter accounts of schools in the UK, this paper provides the first set of large-scale and real-world evidence about how schools have been engaging with and publicly promoting cyber security and online safety education. The findings can not only help evaluate how various cyber security and online safety education initiatives have been perceived by schools but also reveal areas for improvement, e.g., it seems that independent schools have not been sufficiently engaged with cyber security and online safety education initiatives despite their access to more resources.

\iffulledition
In addition to providing valuable insights on cyber security and online safety education, our work can also be easily generalised to study other aspects of cyber security and online safety education and other topics in education. For instance, the semi-automatic verification process of Twitter accounts allows balancing quantity and quality.

Like other similar work on online social media analytics, our work has some limitations, which we hope to address in future work. For instance, we focused on posts from the school's Twitter accounts but ignored replies from other accounts completely. However, capturing conversations between school Twitter accounts and other accounts (e.g., students, teachers and parents) can help see a complete picture from the perspectives of more stakeholders. In addition, using Twitter alone can be biased, so in future, we plan to look at other platforms where schools publicly discuss cyber security and online safety education. Finally, some parts of our data analysis could be further improved, e.g., our engagement diversity analysis is based on 1,000 randomly selected accounts, which may not accurately reflect the whole picture, so conducting a larger-scale analysis will help. Another area for future work is to investigate indirect mentions of organisations via initiatives, which will require a detailed mapping between organisations and initiatives, potentially down to more concrete events of each initiative.
\fi

\bibliographystyle{ACM-Reference-Format}
\bibliography{main}

\end{document}